\documentclass[aps,prl,showpacs,twocolumn,floatfix]{revtex4}
\usepackage{epsfig}
\usepackage{amssymb}
\usepackage{subfigure}
\usepackage{amsfonts}
\usepackage{epsfig}
\usepackage{amsmath}
\usepackage{graphicx}

\newcommand{\be}{\begin{equation}}
\newcommand{\ee}{\end{equation}}
\newcommand{\beq}{\begin{eqnarray}}
\newcommand{\eeq}{\end{eqnarray}}
\newcommand{\beqq}{\begin{eqnarray*}}
\newcommand{\eeqq}{\end{eqnarray*}}

\newcommand{\pp}[2]{\frac{\partial{#1}}{\partial{#2}}}
\newcommand{\eps}{\epsilon}

\def\L{\mathrm{L}}

\def\H{\mathrm{H}}
\def\U{\mathrm{U}}
\def\T{\mathrm{T}}
\def\Gr{\mathrm{St}^{-1}}

\def\Ca{\mathrm{Ca}}

\def\Ha{\mathrm{Ha}}

\begin{document}

\title{Dynamics of Perfectly Wetting Drops under Gravity}
\author{Ryan P. Haskett$^1$, Shomeek Mukhopadhyay$^2$}
\affiliation{$^1$Department of Mathematics, Duke University, Durham, NC 27708; $^2$Department of Physics, Duke University, Durham, NC, 27708}
\date{\today}


\begin{abstract}
We study the dynamics of small droplets of polydimethylsiloxane (PDMS) silicone oil on a vertical, perfectly-wetting, silicon wafer. Interference videomicroscopy allows us to capture the dynamics of these droplets.  We use droplets with a volumes typically ranging from 100 to 500 nanolitres (viscosities from 10 to 1000 centistokes) to understand long time derivations from classical solutions.  Past researchers used one dimensional theory to understand the typical $t^{1/3}$ scaling for the position of the tip of the droplet in time $t$.  We observe this regime in experiment for intermediate times and discover a two-dimensional, similarity solution of the shape of the droplet.  However, at long times our droplets start to move more slowly down the plane than the $t^{1/3}$ scaling suggests and we observe deviations in droplet shape from the similarity solution.  We match experimental data with simulations to show these deviations are consistent with retarded van der Waals forcing which should become significant at the small heights observed.
\end{abstract}

\maketitle

\section{\bf I.Introduction}

The motion of a liquid drop falling due to gravity on a vertical solid substrate is a classic problem in fluid dynamics \cite{1,2}. The movement of the fluid front depends on a balance between viscous and surface tension forces with gravity providing the body force. In the case of extremely small droplets, it is expected that in the late stages of spreading when the drop thins appreciably, subtle interplay between van der Waals and surface tension forces may become important. A detailed study of droplets is also technologically important in applications where the dynamics of droplets with solid substrates is used, e.x. in microfluidic devices and inkjet printing of organic electronic circuits \cite{3}.

\par
When a uniform fluid layer (line source) flows under gravity, the advancing contact line develops a fingering instability which has been the subject of intense study \cite{4,5,6,7,8,9}. In the case where surface tension can be neglected, the contact line shows a typical scaling relationship with time which was analyzed by Huppert \cite{1} for arbitrary angles of inclination.  Since the flow of liquids is technologically important for various coating processes, a through analysis of gravity driven flow for point and line sources (with zero surface tension) was done by Lister \cite{7}.  Recently Gonzalez et al. \cite{8} extended this study to the case of a finite volume fluid strips falling under the action of gravity.  Their careful numerical and experimental work showed deviations from Huppert's similarity solutions due to surface tension effects.  These effects lead to the formation of a capillary ridge in the front before the onset of the fingering instability.  Furthermore Gonzalez et al. predict the wavelength of the fingering instability, noting it can be controlled by the surface tension.  The shape of the profile has been investigated in detail by Hocking \cite{9}, where he identifies three distinct regimes of the fluid sheet before the onset of the instability.

\par
 Depending on the wetting properties of the substrate, the dynamics of finite volume droplets under the combined effects of surface tension and gravity can give rise to a number of interesting shapes \cite{10,11}. The capillary number is the main parameter controlling the morphological shape transitions and dynamics, in contrast to the constant flux flow of a liquid. For droplets with a finite contact angle, one can have corners, cusps or pearl like droplets depending on the capillary number. The corner formation in cusp-like shapes have been linked to the existence of conical similarity solutions in the lubrication equation \cite{12}, while recent numerical work predicts the existence of chaotic shedding states \cite{13}.

\par
In the present work, we concentrate on completely wetting substrate, where the contact angle goes to zero.  This allows us to make detailed comparison between theory, numerical simulations and experiments, without using ``a priori'' contact angle versus disjoining pressure relationship \cite{14}.  We capture the intermediate-time dynamics by a two-dimensional constant volume analysis of the lubrication equation which includes gravity and surface tension.  Only in the long time limit do the effects of the intermolecular forces become important. Careful numerical study of these droplets and qualitative comparison with experiments allows us to understand the importance of disjoining pressure in the dynamics \cite{15}.  We use a precursor film in the numerical simulations \cite{16} and only the long-time dynamics is sensitive to precursor height values.  A comprehensive study of droplets under the completely wetting scenario can reveal how modifications to the disjoining potential have to be made for partially wetting surfaces. It also brings out the important similarities and differences between a constant volume drop and a single finger formed by the instability of the contact line.  Recent work on partially wetting substrates has shown the importance of taking these van der Waals forces into account for very small drops \cite{17}.

\par
In the next section, we give a detailed outline of the constant volume problem, along with the experimental setup used to investigate the problem.  We will also show typical experimental results and how the initial circular shape of the drop changes from a circular plane form to a form with a regular scaling law and then finally to a nearly elliptical plane form.  In Section III, we undertake a detailed theoretical analysis of the two dimensional lubrication approximation for drops with constant volume.  The numerical results are compared to experimental data for position versus time, longitudinal and transverse shape of the drop at intermediate times.  In Section IV, we present numerical results with van der Waals forcing and compare them with experimental results when the dynamics of the front shows strong deviations from $t^{1/3}$ scaling law at very long times.

\section{\bf II. Outline of the Constant Volume Droplet Problem}
\par
We use Polydimethylsiloxane (PDMS) silicone oil as our working fluid on a pretreated silicone wafer.  The oxidized silicon wafer is first wetted with the silicone oil (density $0.986$ g/cc and surface tension $20.9$ dyne/cm), then rinsed thoroughly with hexane and finally cleaned with a methanol wetted tissue \cite{18}.  This covers the substrate with a precursor film of PDMS which is a single monolayer thick and makes it a completely wetting substrate. The PDMS (United Chemical Technologies) oils used had viscosities ranging from $1.5$ centistokes to $1000$ centistokes. Since the timescale for the various effects are controlled mainly by the viscosity, changing the viscosity allows one to follow the change of shape from a circular to an elliptic cross section.  The volume of the drop was varied from around $100$ nanolitres to a few microlitres by using a micropippette (Biohit Proline Micropippetors).

\par
\begin{figure}[tbf]
\centerline{
\setlength{\epsfxsize}{2.0 in} \epsfbox{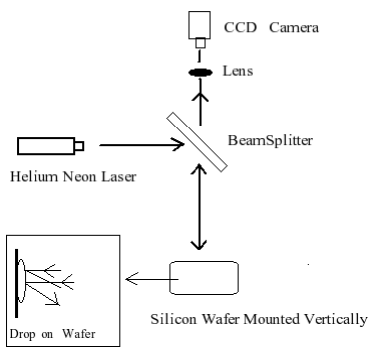}}
\caption{Schematic of the experimental setup where rays reflected from the top and bottom of the drop (inset) creates the interference pattern.}
\label{schematic}
\end{figure}

\begin{figure}[tbf]
\begin{center}
\includegraphics[width=3.5 in]{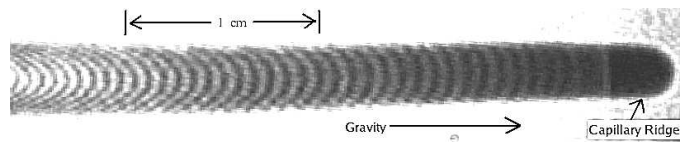}
\caption{ Close up of Interference fringes from PDMS oil droplet of viscosity 100 centistokes and volume 0.85 microlitres. Note the Capillary ridge at the leading edge.}
\label{interf}
\end{center}
\end{figure}

The droplet was placed on the surface of a silicon wafer (which was mounted on a vertical plane) and was illuminated by a collimated He-Ne laser beam (wavelength  $632.8$ nm) as shown in Figure~\ref{schematic}.  The rays reflected from the top surface of the oil drop and the surface of the wafer gave a typical interference pattern as shown in Figure~\ref{interf}.  The position and shape of the front as well as the fringes can be followed as a function of time.  It should be noted that most of the analysis will be concerned with the main body of the drop.  Surface tension effects create a capillary ridge at the tip of the drop.   The back of the droplet moves extremely slowly on timescales compared to the front of the drop.

\par
 From the images (Figure ~\ref{interf}) three main sections of the drop can be distinguished, a dark front which comprises the capillary ridge, the main body of the drop where most of the interference fringes can be seen and the rear section of the drop which moves extremely slowly on the intermediate and long time scales.  In addition to following the front of the drop, the interference fringes (Figure~\ref{interf}) will allow us to draw some qualitative conclusions regarding the surface profile, which can be compared with numerical simulations and theory.

\par
  We note that the height of the fluid is much smaller than the extent of fluid in the plane, and the slope of the free surface is small except near the capillary tip. Hence we can use the lubrication approximation for analyzing the motion of the bulk of the fluid. It is also well known that if the Bond number, $B = \rho g W^{2}/\sigma$, where $W$ is a typical transverse length scale, is much greater than $1$, then the effects of surface tension and contact line motion can be neglected. Depending on the the viscosity and surface tension our typical Bond numbers are around $B \sim 0.2 - 0.8 $. Hence we are in the regime which are in the opposite regime of those analyzed by Huppert and Lister \cite{1,7}.  In the case of a droplet on a partially wetting substrate a function of the Bond number also appears as a control parameter which determines the transition between shedding and nonshedding states \cite{13}.  The capillary lengths for our PDMS oils are around $\sqrt{\sigma / \rho g } \sim  0.1 - 0.2 mm$ where $\sigma$ is the surface tension and $\rho$ is the density of the drop. Assuming a hemispherical drop of volume around $0.2 - 0.5$ microlitres, the initial radius about the same scale as the capillary length.  We choose the free scaling parameter, the height scale, to be this value.

\par
When the drop is initially put on the substrate, high curvature at the contact line pushes the fluid radially outwards. Depending on the viscosity of the drop the circular plan form maintains itself until the gravitational forces overcomes inertial and capillary effects.   If the initial radius of the drop is $R$ and the typical length scale in the downslope direction is given by $L$, then the surface tension driven spreading, which is dependent on the curvature $\kappa$ is given by $\kappa \sim R/ L^{2} $.  As was shown in recent experiments of Biance et al. \cite{19}, in the inertial-capillary regime the length should scale as $ \sim { \sqrt{D t }} $, where the dynamic diffusion coefficient is given by $D = \sqrt{\sigma R / \rho} $.  We do not observe any $\sqrt{t}$ kind of dependence in the timescales when we start taking the first measurements ($1-2$ seconds).  This is borne out by the fact that the dimensionless Ohnesorge number, $Oh = \eta / \sqrt{\sigma \rho R}$, which distinguishes between the inertial and the viscous regime \cite{20}, is around $1$ for the viscosities $\eta$ in our experiments.

\begin{figure}[h]
\begin{center}
\includegraphics [width=3.25 in]{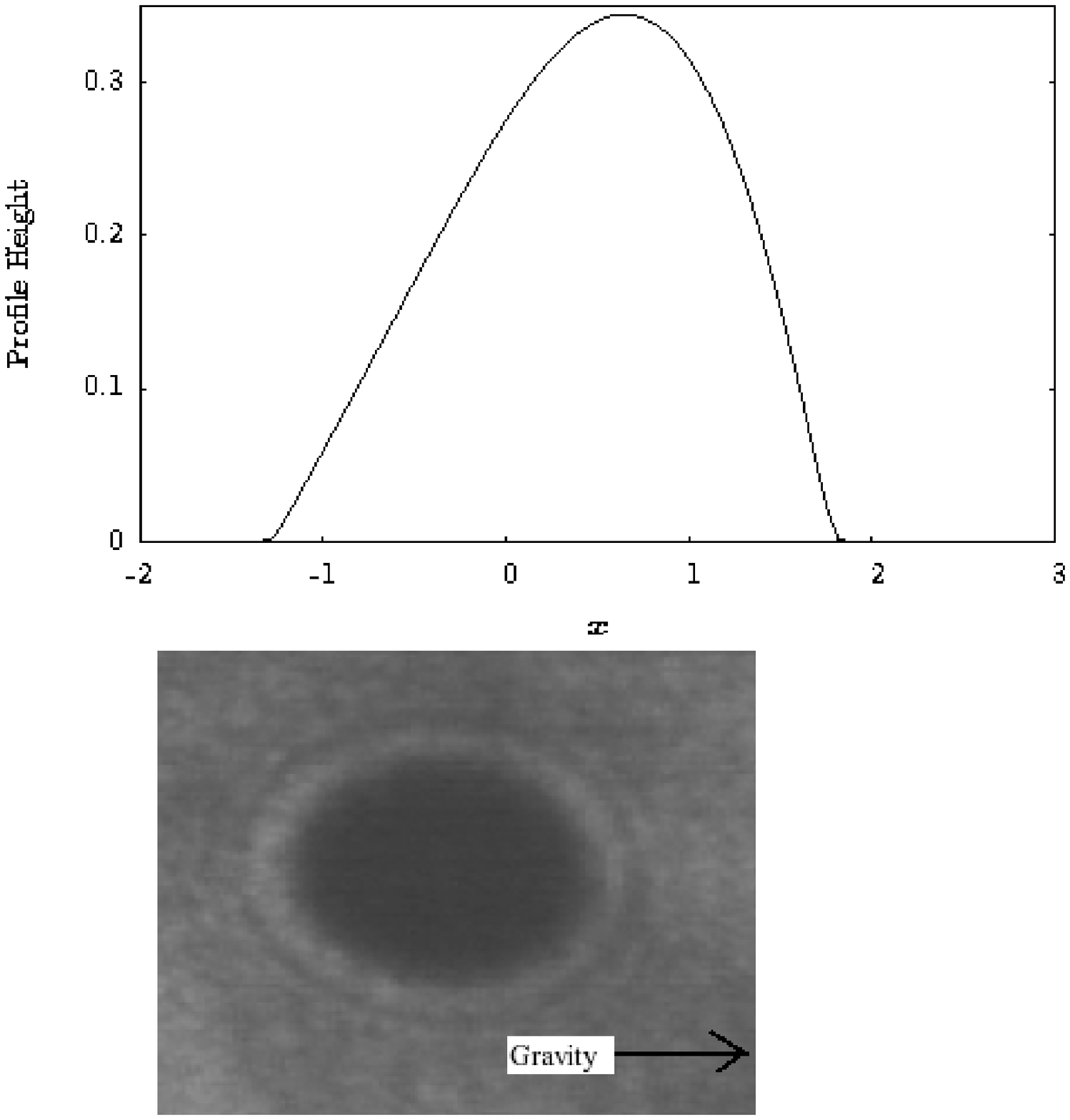} 
 \includegraphics [width=2.5 in]{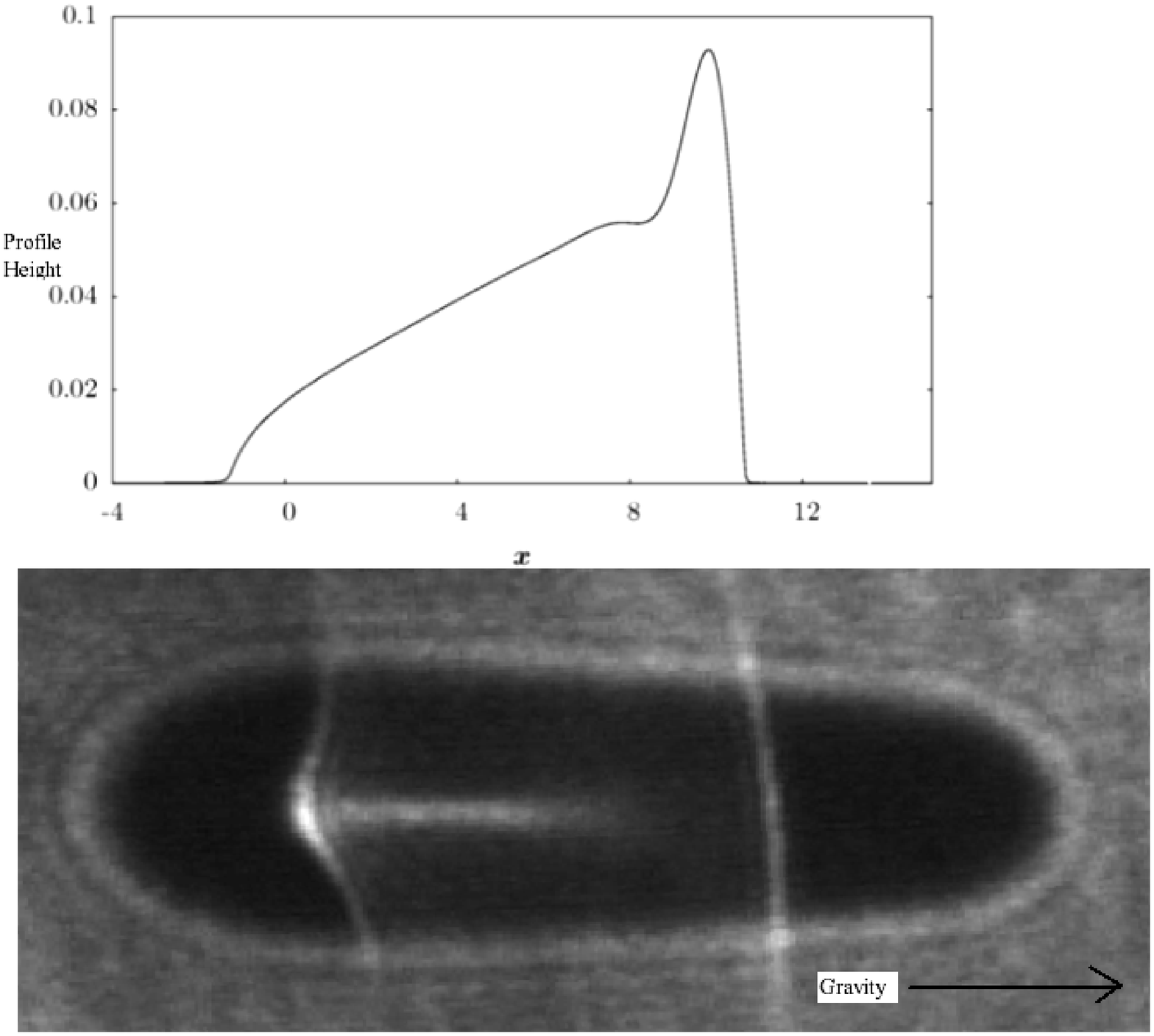}
 \includegraphics[width=2.5 in]{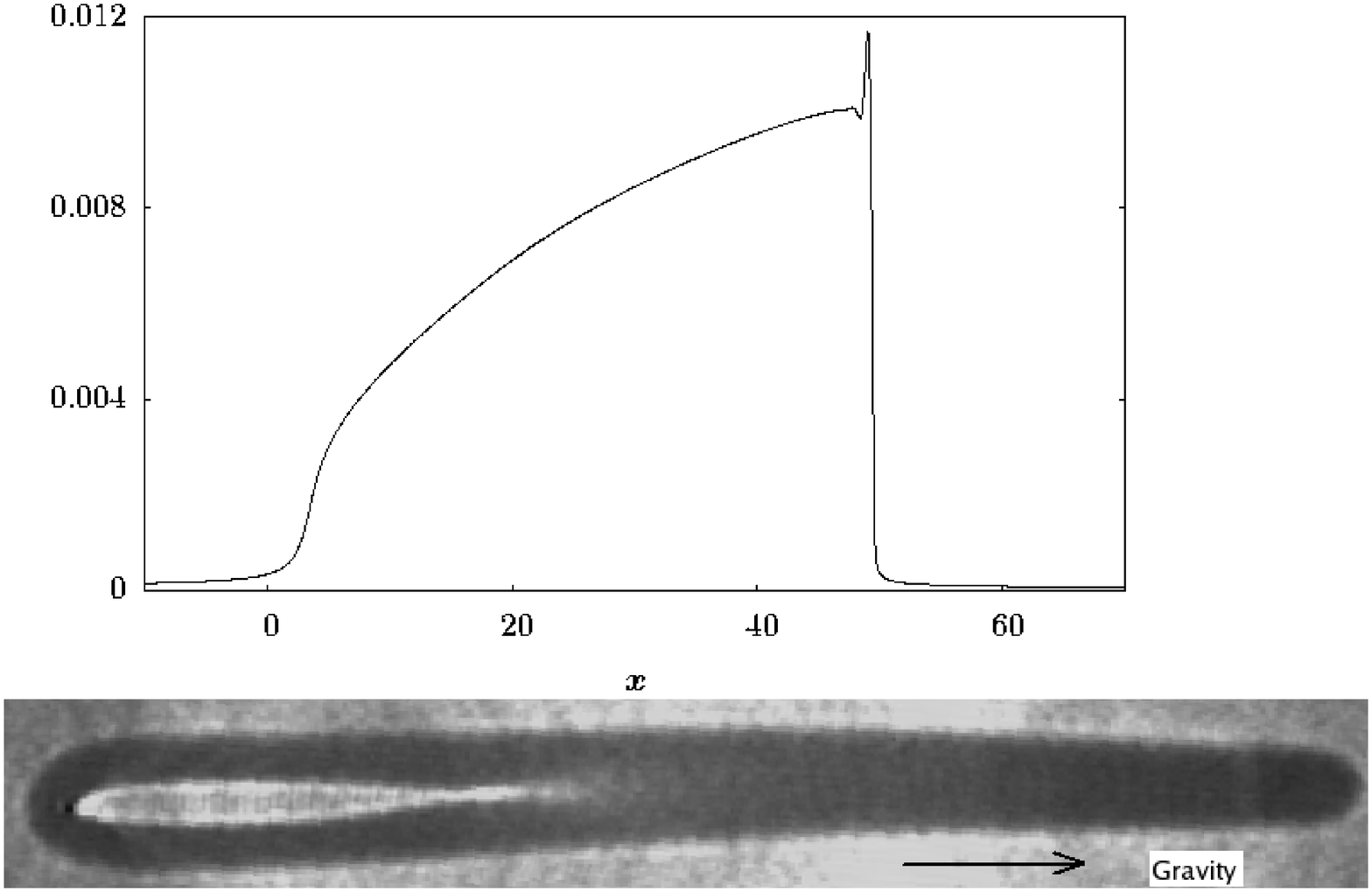}
\caption{The three regimes of droplet motion shown in experimental footprints and corresponding numerical profiles.  The first pictures are from the initial transients.  The second set corresponds to the important asymptotic regime where the gravitational effects are balanced by inertial and surface tension effects.  Finally, as the height scale of the drop becomes much smaller we enter a third regime where Van der Waals forces become important.See section (III) for detailed explanation of numerics.All the numerical heights are in dimensionless units.}
\label{threePic}
\end{center}
\end{figure}

\par
In Figure~\ref{threePic}, we show typical examples from the three main regimes we see in the experiments. Initially, the surface tension forces dominate and we have an nearly circular drop. In the case of a purely gravitationally driven flow, neglecting the effect of surface tension the initial axisymmetric spreading was found to be proportional to $t^{1/2}$ by Lister \cite{7} and gave away to predominantly downslope flow when $t \sim  O(1)$. 

\par
   In the second regime we see a very regular motion where both gravity and surface tension have to be taken into account.  The transients typically die out in the time scale $\approx \sqrt{\eta}$. Looking at the position of the tip as it varies in time, we see $t ^{1/3}$ scaling law relationship (Fig.~\ref{datasets}) for the intermediate times.   Huppert \cite{1} first explained this scaling law using one-dimensional theory without surface tension.  We extend this theory to two dimensions and use asymptotics to allow a detailed understanding of the transverse features of the droplet. 

\par
At very long times, we start to see deviations from the intermediate-time scaling laws.  The motion of the tip slows down significantly, and the droplet starts to spread laterally at an increasing rate.  We note through other works \cite{11,17} that at very small height ranges near the end of our runs, van der Waals forces should start to become important.  Looking at the effects of these forces in our numerics we can see the same shape changes as we see in the experiments.  In particular, we can match the deviation in tip position from the $t^{1/3}$ scaling law in experiment and simulations.  From this evidence we conclude that van der Waals forces are the cause of long-time deviations from the classical solutions seen at intermediate times.

\section{\bf III Formulation}
\par
\begin{figure}[tbf]
\centerline{
\setlength{\epsfxsize}{3 in} \epsfbox{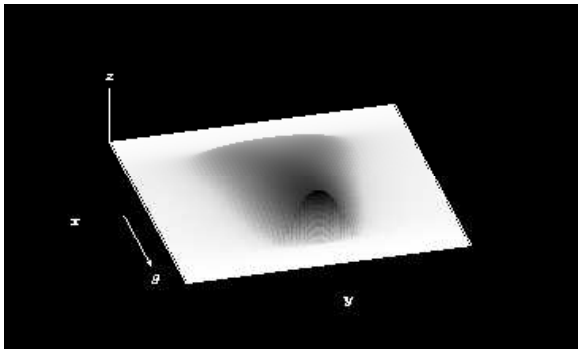}}
\caption{Geometry of the problem}
\label{Geometry}
\end{figure}

We study the gravity-driven lubrication flow of a small droplet of Newtonian fluid on a vertical solid flat plate. The finite-mass droplet has a constant density $\rho$, viscosity $\mu$ and surface tension $\sigma$, and the flow is driven by gravitational stresses and interacts with the fully wetting surface.  We define a coordinate system with the $x$-coordinate in the direction of gravity and the $z$-direction normal to the plate, as shown in Figure~\ref{Geometry}.  The $y$-direction is transverse to the direction of flow and completes the right-handed coordinate system.  We denote the height of the free-surface of the flow as $z=h(x,y,t)$.

\par
  We shall now briefly review the derivation of the lubrication model including long range molecular forces \cite{21}. We rescale the dimensional (hatted) lengths, time, velocity, and pressure to yield corresponding dimensionless quantities,
\begin{equation*}
\begin{split}
 \label{rescaling}
	&\hat{x}	= \L\, x, \qquad
	\hat{y} = \L \, y, \qquad
	\hat{z}	= \H\, z,	\qquad
 	\hat{t}	= {\L\over \U}\,  t, \\
	&\hat{u}	= \U\, u,	\qquad
	\hat{v} = \U \, v,	\qquad
	\hat{w}	= {\H\U\over\L}\, w, \qquad
	\hat{p}	= {\mu \U \L\over \H^2}\, p.
\end{split}
\end{equation*}
 and use these rescalings on the the Navier-Stokes equations for the viscous, incompressible flow where the modified pressure $\bar{P}$ directly incorporates the hydrostatic pressure and the van der Waals molecular interactions,
\begin{equation}
	\bar{P} = p + x\, \Gr + \Ha \,\Pi (h). 
\end{equation}
The dimensionless parameters present are the aspect ratio, the Stokes and Hamaker constants as well as the reduced capillary number,
\begin{equation*}
	\eps	= {\H \over \L},
	\Gr 	= {\rho g \H^2 T\over \mu \L},
	\Ha	= {A' \T \over 6 \pi \mu \H^2 \L^2},
	\Ca	= {\eps^{-3}\mu\U \over \gamma}
\end{equation*}
  where $g$ is the acceleration due to gravity and $A'$ is the dimensional Hamaker constant.

\par
At $z=0$, the no-slip boundary conditions, $u=w=0$, are applied.  At the free-surface, $z=h(x,t)$, the boundary conditions are the kinematic stress balance condition, the normal stress balance and the tangential stress balance.  We follow the derivation in Oron, Davis and Bankoff \cite{21} integrating the Navier-Stokes equations to find the depth averaged velocity and then using the kinetic boundary condition to produce a partial differential equation for the height,
\begin{multline}
	h_t + \frac{\Gr}{3} (h^3)_x \\
		=  \Ha \nabla \cdot \left(h^3 \nabla \Pi(h) \right) - \frac{\Ca ^{-1}}{3} \nabla \cdot (h^3\nabla \nabla^2 h).
\label{GE1}
\end{multline}
 where $\nabla = (\pp{}{x},\pp{}{y})$ is the gradient on the plane. 

\par
Despite the relatively small constant in front of the van der Waals term in (\ref{GE1}), we must keep this term to understand the longtime dynamics, as we will show in section IV. We will use $1/h^4$ form for the van der Waals force $\Pi(h)$, which comes from the dielectric properties present in the liquid-solid layered system when the liquid has no ionic species \cite{22}.  We choose to incorporate the aspect ratio into the Hamaker constant because the van der Waals interactions are important near the contact line at the edge of the droplets.  This term then must be included to correctly model the physics despite the small value of the Hamaker constant in the lubrication limit ($\eps \rightarrow 0$).  As the film thins the `retarded' or long-range part of the van der Waals interaction becomes important.  Casimir and Polder \cite{23} first showed that in the limit of long distances the interaction energy for the van der Waals interaction changes from $ 1/r^6$ to $ ~ 1/ r^7$.In order to calculate the disjoining pressure from this functional form one has to integrate the two parallel domains using the full Lifshitz theory, which has been done by Derjauin et al. \cite{22}. This `functional' form of the van der Waals interaction comes into effect when the film thickness is above 10 nanometres. The retarded van der Waals interaction has been used in modelling the dynamics of thin tear films in the eye amongst others \cite{24,25}.  For the disjoining pressure of the form $A'/h^4$, $A'$ has dimensions of erg-cm, whereas for the non-retarded form, ${ A'/h^3}$, $A$ has dimensions of ergs.

\par
We now choose our scaling similar to \cite{26}, 
\begin{equation}
	\L = \left(\frac{\sigma H}{\rho g } \right)^{1/3}, \qquad
	\T = \frac{3 \mu L}{\rho g H^2 },
\end{equation}
which sets all the coefficients in front of the terms in (\ref{GE1}) to unity except for the Van derWaals interaction term,
\begin{equation} 
\label{GovEqn}
 	\pp{h}{t} +  \pp{h^3}{x} = - A \nabla \cdot \left( h^3\nabla \frac{1}{h^4} \right) 
 		- \nabla \cdot (h^3 \nabla \nabla^2 h).
\end{equation}
We keep the ability to adjust or change the strength of the van der Waals interactions though the non-dimensional parameter,
\begin{equation}
	A = \frac{A'}{6 \pi} \left( \frac{1}{\sigma \rho^2 g^2  H^{13}} \right)^{1/3}
\end{equation}
so we can understand the effect of the intermolecular forces on the fluid.  Our other major parameter in the problem will be the volume of the droplet $V'$ which when scaled becomes $V= \frac{V'}{H L^2}$.  However, for our comparisons between numerics and experiments we can avoid dealing with this parameter by setting our free height scale $H = \frac{L^2}{V'}$, and therefore obtaining a scaled volume of $\sim 1 $.

\par
One-dimensional numerical calculations were preformed using standard Crank-Nicolson framework \cite{27}.  All two-dimensional numerical solutions for the governing equation (\ref{GovEqn}) and the scaled version of this equation we discuss later (\ref{SGE}) use a midpoint, alternating direction scheme as seen in \cite{28}.  These methods along with Newton's iterations for the non-linearity, allow for second-order accuracy in time and space while keeping reasonable execution time.

\par
 In all cases, we must estimate the pre-wetting height of film used in the experiments as this value is not directly measurable.  This was chosen to be two to three orders of magnitude less then the smallest heights achieved by the droplet.  Throughout this range we find the same qualitative results seen in the next sections for scaling powers and droplet shapes before the van der Waals forces become important.  As the film height becomes extremely small, however, the strength of the van der Waals term is affected by the wetting height.  This can be understood by looking at the terms in the governing equation (\ref{GovEqn}), and noticing that most of the terms depend on positive powers of $h$ while the van der Waals has a negative power of $h$ and therefore is more sensitive to small changes in $h$.  Pre-wetting dependence for the quantitative motion of fluid has been hypothesized by many authors \cite{4,8,21} as explanations for large deviation in proportionality constants while still having very good fits for scaling powers.  We clarify the cause of these deviations in the Appendix.

\subsection{\bf A One-Dimensional Solutions without van der Waals Forcing}
\par
  We start the analysis of one dimensional solutions without van der Waals forcing by taking the governing equation (\ref{GovEqn}) in one-dimension, for solutions that are uniform in the $y$-direction and with no wetting i.e. $A=0$ we obtain the governing equation
\begin{equation} 
\label{GE1D}
 	\pp{h}{t} +  \pp{h^3}{x} = - (h^3 h_{xxx})_x.
\end{equation}
  This equation has been studied extensively for both constant volume \cite{1,8} and constant flux \cite{7} cases and we will review the main results here.   We verify our results in intermediate times before van der Waals forces become important using theoretical results from this equation.  Also, we use the scaling laws presented to understand the novel, two-dimensional asymptotic solution in the next section.

 Ignoring the fourth order smoothing for the moment, we obtain the same equation that Huppert first used \cite{1} to describe the motion of a droplet,
\begin{equation}
	\pp{h}{t} +  \pp{h^3}{x} = 0.
\end{equation}
 with the solution
\begin{equation}
\label{phi0}
	h_{A=0}(x,t) = \sqrt{\frac{x}{3t}}
\end{equation}
  valid from an arbitrarily chosen $x=0$ to the tip at $x_{\mathrm{tip}} = (\frac{3 \sqrt{3}}{2} V_A)^{2/3} t^{1/3}$ determined by conserving the profile area $V_A$.  The solution is zero for $x$ outside this region.  The appearance of a ``shock'' or ``jump''is not too surprising as we have removed the fourth-order smoothing.  However, the main effect of this smoothing is to produce a capillary ridge, which appears due to the finite surface tension of the fluid, that smooths out the shock at $x_{\mathrm{tip}}$.

\par
\begin{figure}[h]
\begin{center}
\includegraphics[width=2.5 in]{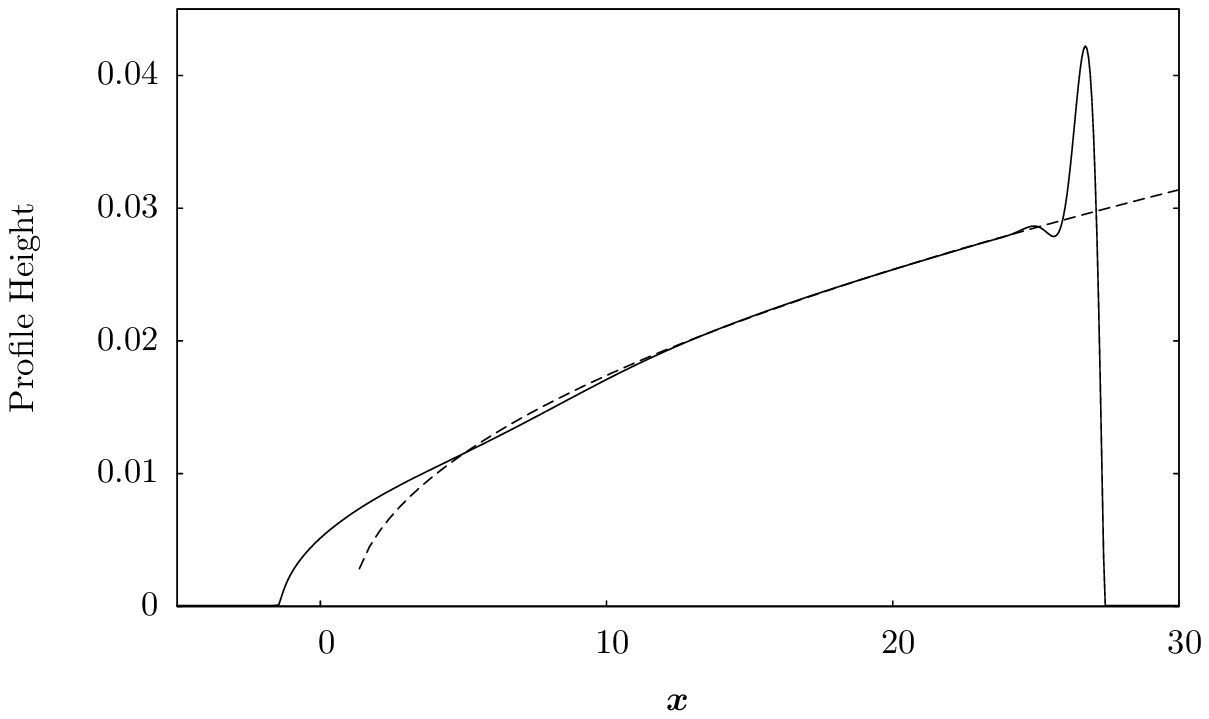}
\includegraphics[width=2.5 in]{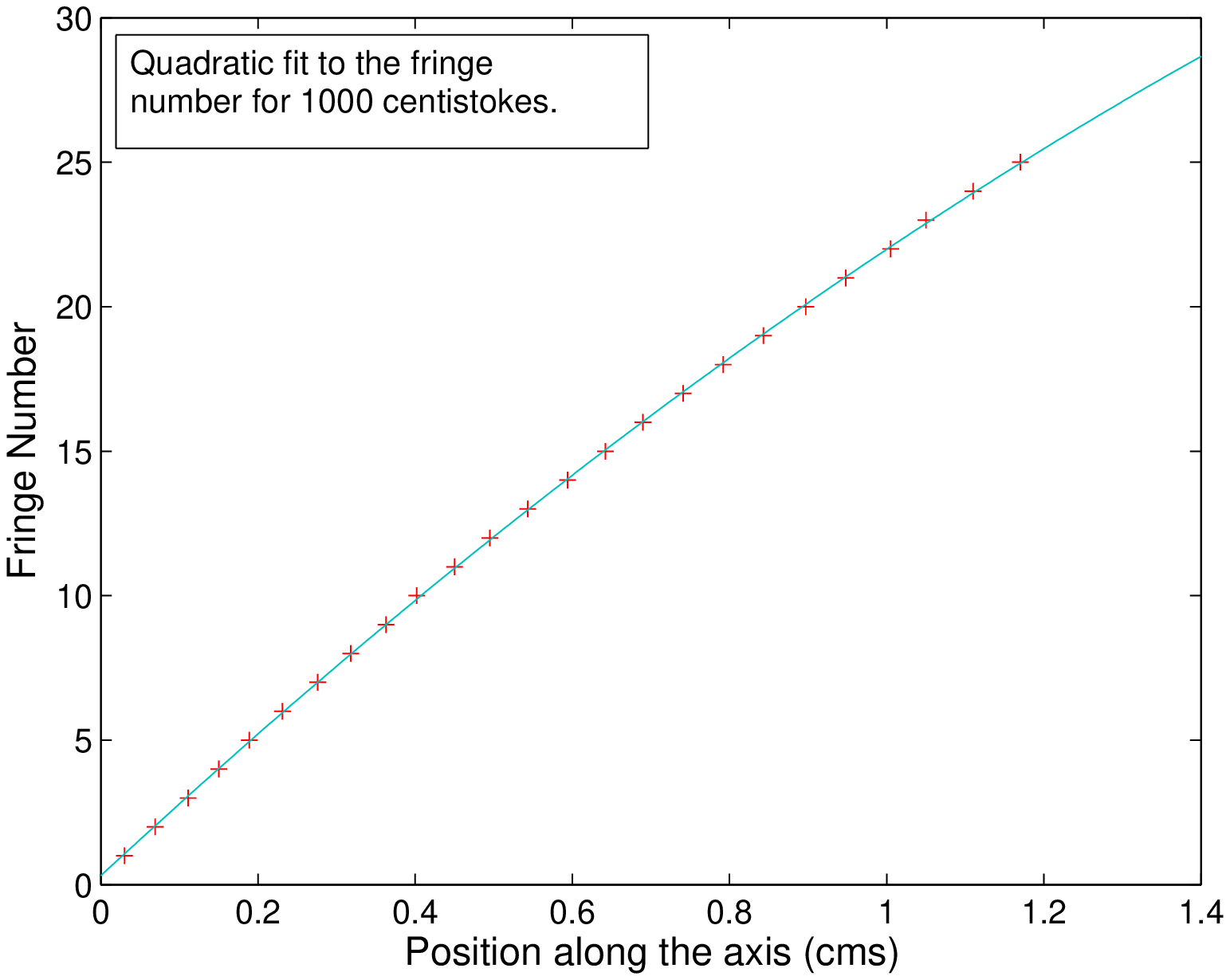}
\caption{A numerical profile $h(x,y=0)$ fit with $C x^{1/2}$ and the analysis of the interference fringes from a $1000$ centistokes run.Fringe profile is fit to a quadratic function in the intermediate section of the drop away from the tip(shown for the first 25 fringes).}
\label{profileFig}
\end{center}
\end{figure}

  The numerics show the predicted square root scaling (\ref{phi0}) of the profile height (Fig.~\ref{profileFig}), which can be compared with the experimental data from analyzing the spacings in interference fringes.  This shape of the the profile is markedly different from the profile in the case of gravity driven fingers from a line source \cite{26}, or in the case of Marangoni driven fingering \cite{18}.  For Marangoni driven fingers or for gravitational fingering from a line source, one has a very flat intermediate section followed by a sharp tip at the end.

\par
\begin{figure}[h]
\begin{center}
\includegraphics [width= 3.5 in]{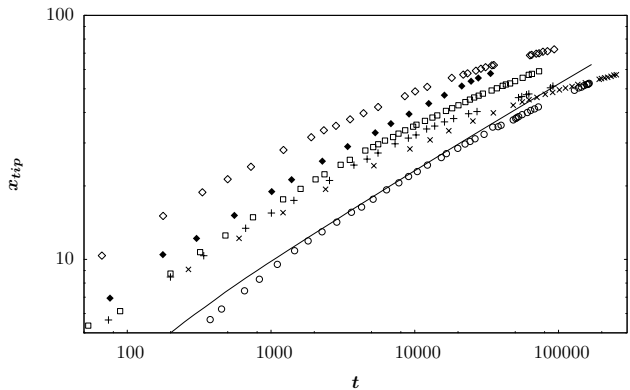}
\caption{Typical data sets for viscosities of $100$ or $50$ centistokes and volume ranging from $100$ nanolitres to $1.0$ microliters graphed with a reference numerical solution (solid)for $A=0$ showing the $t^{1/3}$ scaling. Squares and diamonds correspond to 50 centistokes, others are 100 centistokes. Note the $t^{1/3}$ scaling in the data at early times and long time deviations from this scaling in all cases.}
\label{datasets}
\end{center}
\end{figure}

  The scaling law for the tip position $x_{\mathrm{tip}} \propto t^{1/3}$ has been seen throughout the literature for gravitationally driven fluids \cite{1,4}.  In our experiments and numerics in one and two dimensions, we see this scaling for intermediate times. In Figure~\ref{datasets}, we show typical experimental results for 50 and 100 centistokes of assorted volumes along with a numerical run with $A=0$,to show the regime of $t^{1/3}$ behaviour .  The numerical run has, as expected, a nearly perfect $t^{1/3}$ scaling, however, we see deviations from this scaling in the experiments at long times which will be attributed to van der Waals forces in later sections.

\par

\subsection{\bf B Two-Dimensional Solutions without van der Waals Forces}
\begin{figure}[h]
\begin{center}
\includegraphics[width=2.5 in]{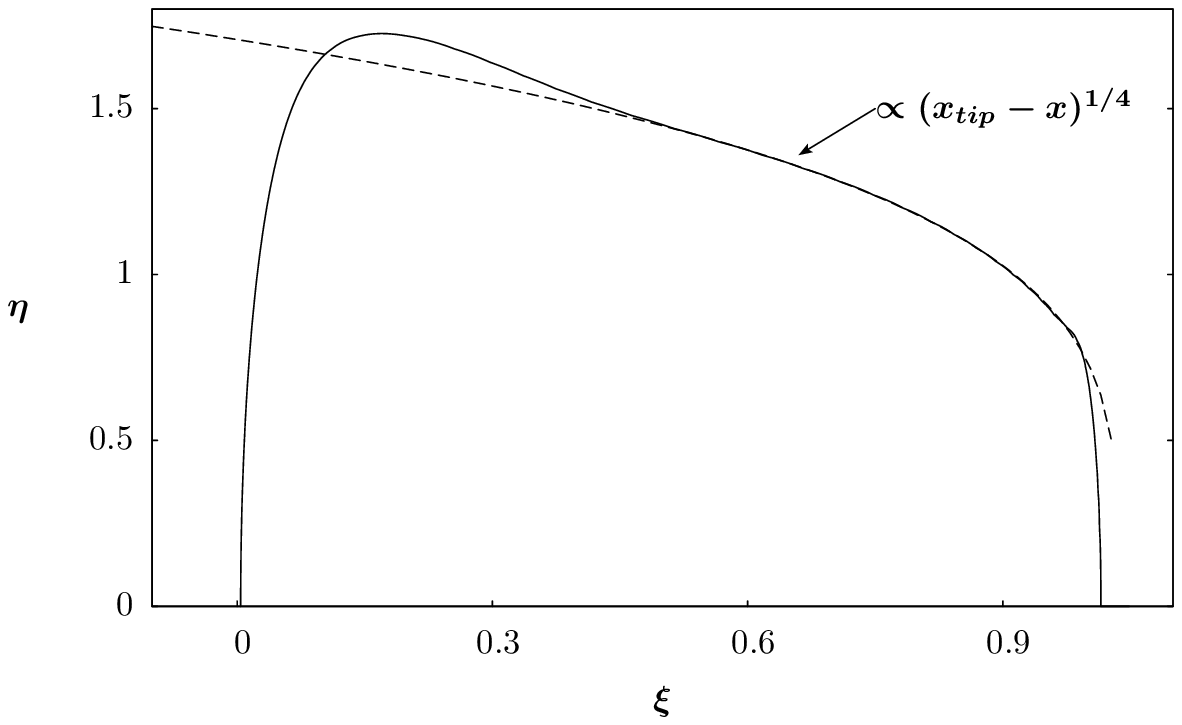}
\includegraphics[width= 2.5 in]{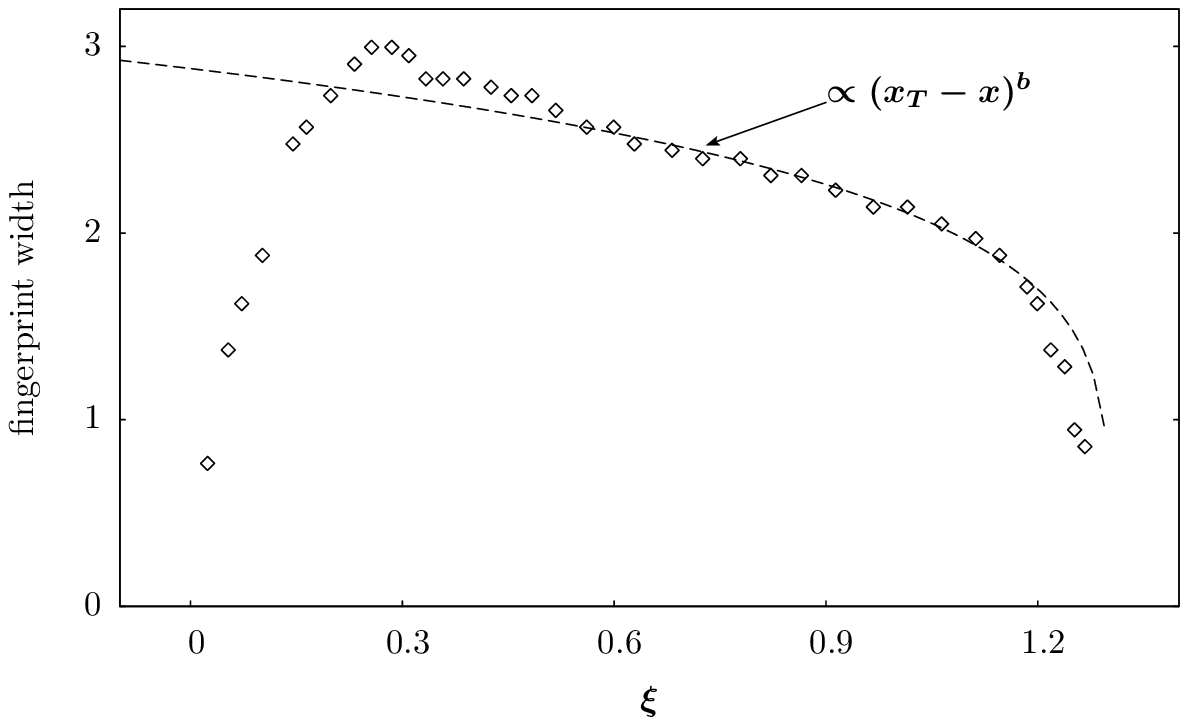}
\caption{The top half ($\eta>0$) of a numerical footprint from a solution of (\ref{GovEqn}) under scaling (\ref{SimScale}) fit with a scaling constant of $1/4$th (top) and an representative experimental footprint (volume $0.1 \mu l$, viscosity $100$ cSt) fit to a scaling function.  As with all of the experimental runs for an intermediate times where van der Waals forces are not yet important.  The scaling constant (in this case $b=0.23$) was close to the predicted value ($1/4$).}
\label{footprint}
\end{center}
\end{figure}

 In the Appendix, we derive a two-dimensional, long-time, asymptotic (near the tip), series solution to our governing equation (\ref{GovEqn}) 
\begin{equation}
\label{AnalSol}
	h(x,y,t) = \sqrt{\frac{x}{3t}} \left( 1 - \frac{y^2}{w(x,t)^2} \right) + \mathcal{O}(y^6)
\end{equation}
where $w(x,t) = W ( 1 - x t^{-1/3}/\xi_{\mathrm{tip}})^{1/4}$ is the width of the droplet.  Note, our solution still has two constants that need to be fit $W$ and $\xi_{\mathrm{tip}}$, due to the lack of a matching asymptotic solution for the back of the drop (near $x=0$).  The first constant $W$ controls how quickly the footprint width scales with the distance from the tip.  The second constant $\xi_{\mathrm{tip}}$, in the physical variables, becomes the scaling constant for the tip position $x = \xi_{\mathrm{tip}} V^{8/15} t^{1/3}$.  Numerically, we have noticed that these constants are quite independent of initial droplet shape as long as the initial volume is contained within an area that is about half of the eventual width $W$.  However, as we look at the experimental values for $\xi_{\mathrm{tip}}$ from Figure~\ref{datasets} and $W$ from footprint data in Figure~\ref{footprint}.  We find quite a bit of spread in these constants despite the very close agreement for the scaling.  This spread in the values has been seen in other droplet literature \cite{1,4} and is discussed in the Appendix.The asymptotic solution is not valid near the tip of the droplet.  The center line height ($y=0$) matches the square root scaling discussed earlier in the one-dimensional theory (\ref{phi0}).  Also, if we look for the point where the width of the drop $w(x,t)$ becomes zero (the tip of the droplet), we see clearly the tip scaling $x_{\mathrm{tip}} \propto t^{1/3}$, as discussed in the previous section. In addition to matching the expected one-dimensional theory, we have a novel explanation for a number of the transverse features we see in experiment.

  We clearly see the inverse parabolic cross-sections predicted by (\ref{AnalSol}) (for constant $x$ and $t$) in the numerics as well as the experiments (Fig.~\ref{parabola}), when data is taken away from the tips of the droplet.  In the Appendix, we show that the $O(y^4)$ term in the series solution must be zero therefore our parabolic fits are nearly perfect.  Experimentally, these parabolic cross-sections are seen for intermediate times before the van der Waals term becomes important.
\begin{figure}[h]
\begin{center}
\includegraphics[width= 2 in]{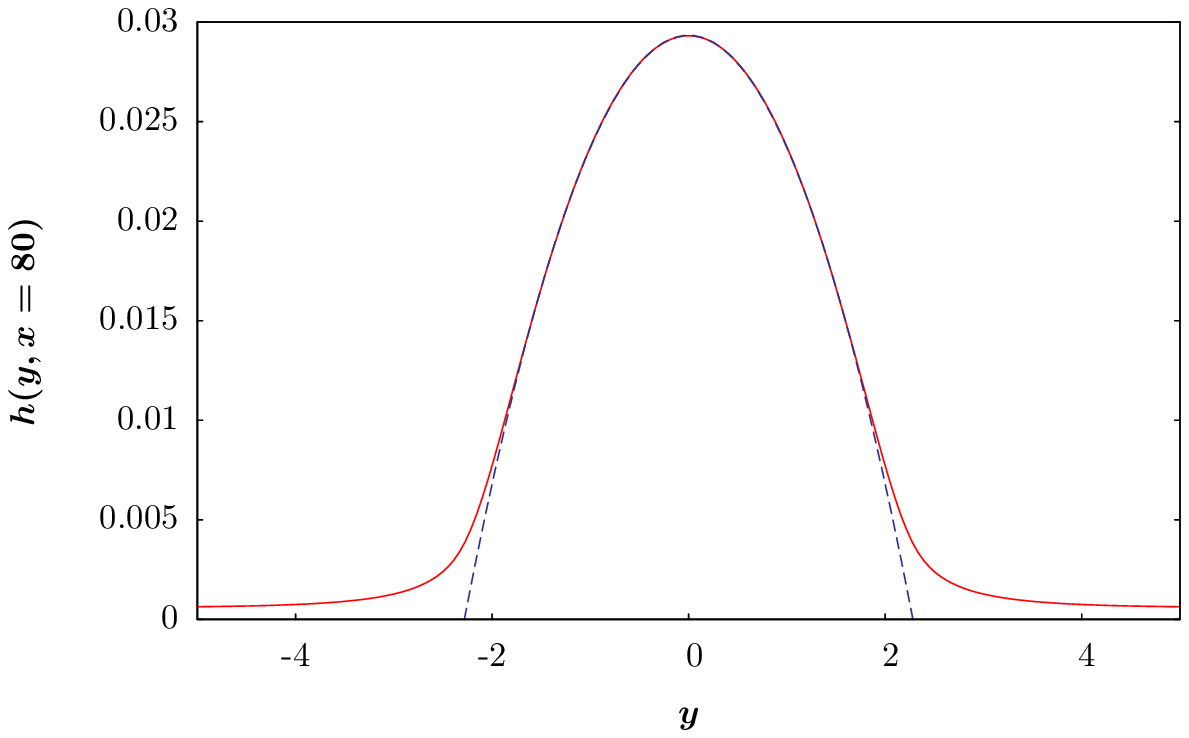}
\includegraphics[width =2 in]{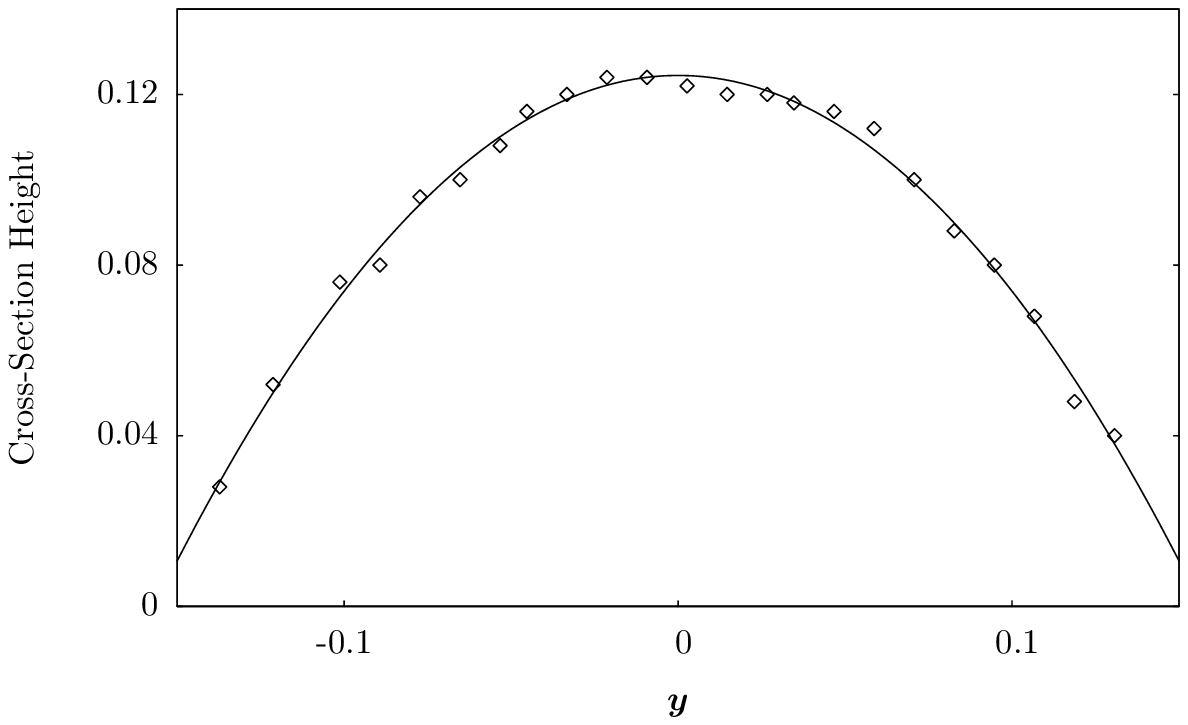}
\caption{A typical numerical cross-section (taken at an $x$ value of about 80\% of the total drop length) for a solution to (\ref{GovEqn}) with $A=0$ (top) and a typical experimental cross-section (taken at a postion about 80\% of the total drop length) at intermediate times fit to a parabola}
\label{parabola}
\end{center}
\end{figure}
  The final important feature we can predict using the asymptotic solution (\ref{AnalSol}), is the shape of the footprint of the droplet (where the height becomes zero).  The theory predicts a $1/4$th scaling law for the width $w$ of the droplet as a function of position along the droplet $x$.  Our numerical solutions for $A=0$ show the correct $1/4 ^{th}$ scaling in the width of the droplet in the footprint, estimated here by a contour of height $h = 0.05$ (Fig.~\ref{footprint}).  Note that we see some deviation at the tips of the droplets.  The deviation for $x$ near zero is not surprising as we derived our solution asymptotically near the tip of the drop. However, we see that the solution is surprisingly valid through most of the droplet length which again can probably be attributed to the zero $O(y^4)$ term.  Also, in the appendix we have scaled out the $x$ derivatives near the tip which should become relatively small over time.  However, as our numerics are not steady st
 ate solutions, but merely long time solutions, and as the $x$-derivatives remain strong right near the edge of the drop, we see the $1/4$th power law cuts off just before the tip.

\section{\bf IV The Effects of van der Waals Forcing}

  In this section we will analyze the effects of the van der Waals term on the shape of the droplet. The relative strengths and applications of the `retarded' and `non-retarded' Hamakar constant has been discussed in detail by French \cite{29}.  When the completely wetting droplet thins, the tip can typically reach a height of less than $100$ nm, which is typically in the range where retarded Hamaker constant will be effective.  In fact, for horizontal droplets ellipsometric measurements during the late stage of spreading show an advancing `precursor' layer which is about $30$ nm in thickness \cite{16}. In recent experiments on spreading of PDMS on partially wetting substrates, for drop volumes of about 100 nanolitres there was a crossover from surface tension to van der Waals spreading \cite{17}. Extensive work on crossover from `non-retarded' to `retarded' forces for liquids on quartz systems have been analyzed by Teletzke et al. \cite{30}.
\par
\begin{figure}[h]
\begin{center}
\includegraphics[width= 2.5in]{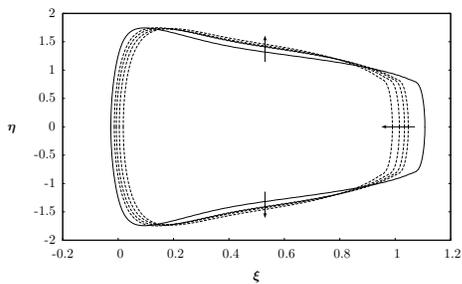}
\caption{Scaled numerical footprints of the droplet at various long times for $A= 4 \times 10^{-12}$.  Note the solution starts near the $A=0$ similarity solution (solid) at intermediate times and we can see significant shape change at increasingly long times (in the direction of the arrows) due to van der Waals forces.}
\label{Afootprint}
\end{center}
\end{figure}

    To understand the effects of the van der Waals forcing we turn to a change of variables suggested by the scaling in the tip position $x \sim t^{1/3}$. Since our motion is mostly one-dimensional at long times, we choose to scale our spatial variable $\xi$ to keep the tip position constant.  This change of variables is a modified version of a scaling proposed by Hocking \cite{31}, from which we can numerically find a similarity solution to our problem (see Appendix) without the van der Waals forcing.  By moving into the scaling variables from the similarity solution we can easily observe the changes caused by the van der Waals term  We then pick the rest of our variables, $\eta$ the transverse scale, $s$ as our ``long'' time variable and $\phi$ the scaled height, to asymptotically balance the governing equation in $t$.

\beq
\label{SimScale}
\begin{split}
	\xi 	&= t^{-1/3} x \\
	\eta	&= y \\
	s	&= \ln{(t)} \\
	\phi(\xi,\eta,s)	&= t^{1/3} h(x,y,t)
\end{split}
\eeq
  We apply this change of variables to the full governing equation with wetting $0 < A \ll 1$ and obtain the following partial differential equation for the scaled height $\phi$
\begin{multline}
\label{SGE}
	\phi_s + (\phi^3 - \xi \phi/3)_\xi = A e^{4s/3} \left( \frac{\phi_\eta}{\phi} \right)_\eta + A e^{2s/3} \left( \frac{\phi_\xi}{\phi} \right)_\xi \\
	- (\phi^3 \phi_{\eta \eta \eta})_\eta - \mathcal{O}(e^{-2s/3}).
\end{multline}
For very small $A$ our solution will evolve into to the two-dimensional solution (\ref{AnalSol}) for intermediate times ($A \exp{(4s/3)} = A t^{4/3} \ll 1$), but it is clear from the scaling that the van der Waals forces will eventually become dominant.  Importantly, the van der Waals forcing in the transverse direction becomes important much earlier $t \approx A^{-3/4}$ while the $x$-direction term kicks in at $t \approx A^{-3/2}$.  From this scaling difference we expect to observe transverse spreading in the experiments.  Note the numerical footprints in Figure~\ref{Afootprint} spread to approximately the footprint of solution with $A=0$ but then continue to spread in the transverse direction over longer times.  We see the same qualitative changes in the experimental footprint (Fig.~\ref{spreadData}) as well as a loss of the $1/4$th scaling of the width predicted in the Appendix. 

\begin{figure}[tbf]
\begin{center}
\includegraphics[width= 2.5in]{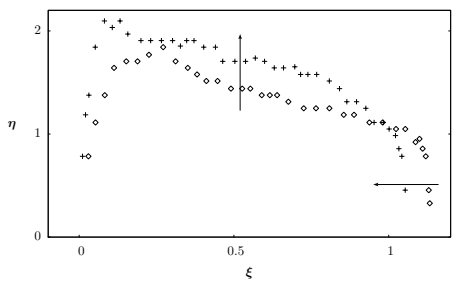}
\caption{Scaled experimental footprints ($100$ centistokes) at an intermediate time ($1550$ seconds, diamonds) and a relatively late time ($14500$ seconds, crosses) showing qualitatively the same transverse spread seen in the above numerical runs.}
\label{spreadData}
\end{center}
\end{figure}

\begin{figure}[tbf]
\begin{center}
\includegraphics[width=2.5  in]{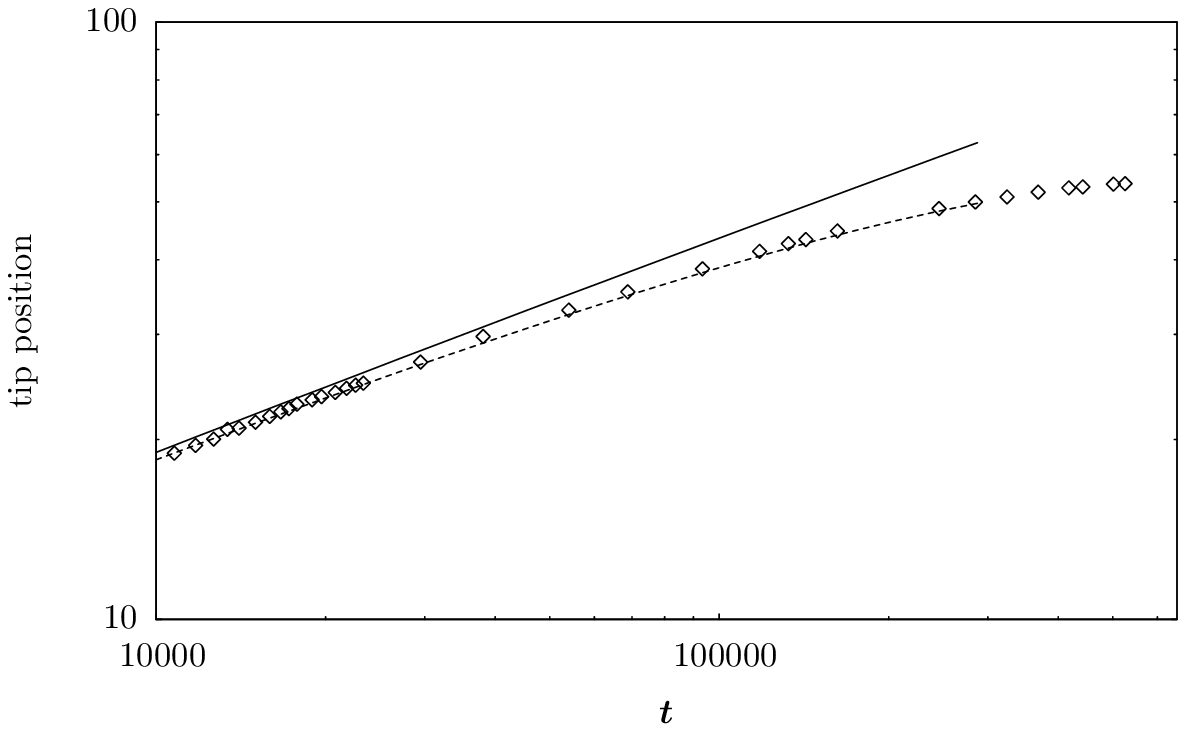}
\includegraphics[width=2.5 in]{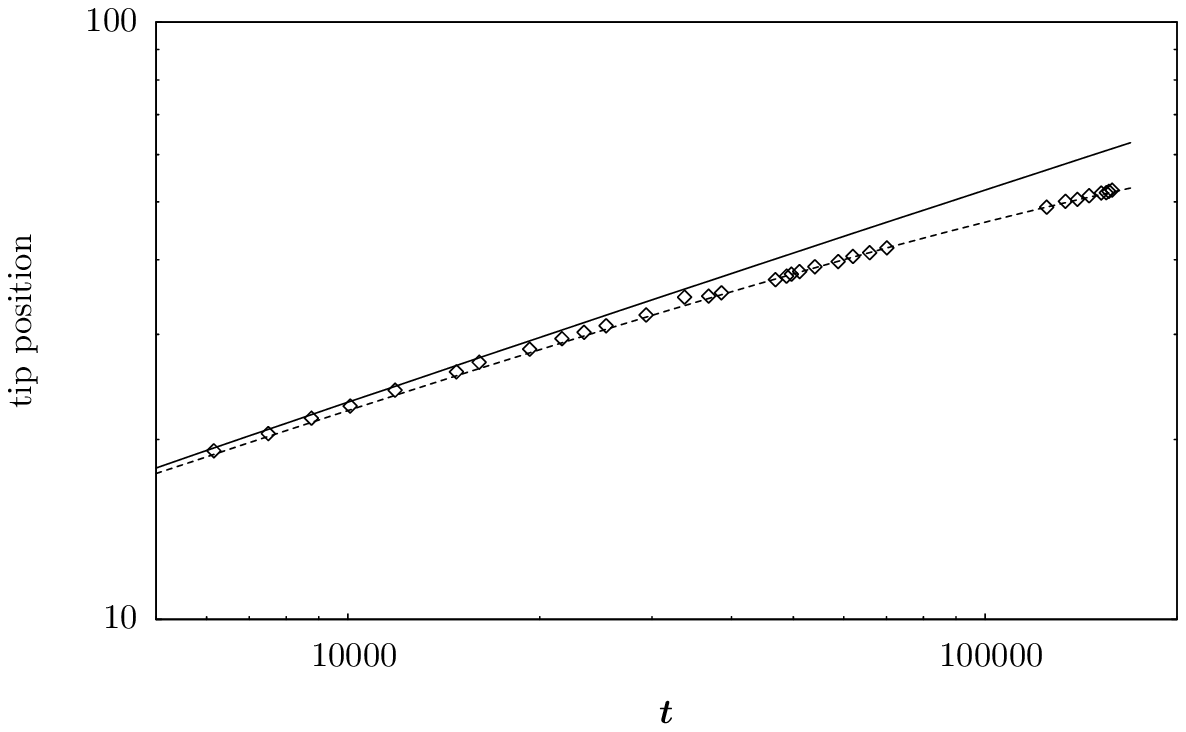}
\caption{A data set with viscosity $50$ centistokes (above)graphed with the closest fit numerical solution $A = 2 \times 10^{-12} \pm 1 \times 10^{-12}$ and wetting height $6 \times 10^{-5}$.A similar data set with viscosity $100$ centistokes (below).}
\label{datafit}
\end{center}
\end{figure}

 The lateral spreading causes deviations in the shape of the drop from the similarity solution without van der Waals forces (\ref{AnalSol}).  Through numerical solutions we observe the transverse spreading slightly reduces the height of the profile height from the values predicted by the similarity solution.  We also see the cross-sections are no longer parabolic to $\lvert \mathcal{O}(y^6) \rvert $.  These deviations are too small to observe in the experiments.  However, we can easily observe a reduction in the speed at which the tip of the drop moves down the plane in the unscaled variables as shown in (Fig.~\ref{datafit}).  These deviations from the similarity solution start at the correct time scale $A \exp{(4s/3)} = A t^{4/3} \approx 1$.  We can fit various data sets and compared their deviation in the tip position from the scaling law $x_t \approx t^{1/3}$.  Using numerics with various values of $A$ we can find an approximate strength of the van der Waals forcing for that experimental run as shown in Figure~\ref{datafit}.

\par
   However, in obtaining this value for the dimensionless Hamaker constant $A$ we had to make many assumptions that could significantly effect the final value.  Some complications arise in fitting stem from the undetermined constants in the solution (\ref{AnalSol}).  As initial lateral spreading and mass distribution can vary within the experiment, we choose to fit the data using an effective volume instead of the measured volume.  The effects of the wetting height (which is unmeasurable), are even more significant as small changes in the wetting height can have fairly large effects on the deviation seen and therefore the matching value for $A$.  Also, the deviation is similar for different functional forms for the van der Waals pressure $\Pi(h)$.  For the forms of the pressure tested in the numerics ($\Pi(h) = 1/h^3$, $\Pi(h) = 1/h^4$, $\Pi(h) = C_1/h^3 + C_2/h^4$) we see very similar qualitative results for the deviation from the $1/3$rd scaling law. In particular, we find that the van der Waals forces are the contributing factor for the long time deviations from the expected $t^{1/3}$ scaling law for the tip position.

\section{\bf V Summary and Conclusion}

As noted earlier, the scaling exponents fit very well for intermediate times, however, in our experiments as well as the experiments of others the constants associated with these scaling laws vary rather widely.  Other authors \cite{4,8,21} have hypothesized that wetting effects contribute to this spread in values.  From (\ref{phiSolution}) in the Appendix, we can see a possible method for wetting effects to cause this spread.  A matching asymptotic solution near the back of the drop ($\xi=0$) would almost certainly depend on the form and strength of the wetting, as the height is very small even at intermediate times.  This matched solution would determine the value of $W$ and therefore (from conservation of mass), the  $\xi_{\mathrm{tip}}$.  So, the van der Waals form could well affect the scaling constants strongly even though they don't explicitly affect the droplet shape for intermediate times.
\par
  The experimental and numerical evidence clearly points to van der Waals forces as the cause of long time deviations from classical solutions for the motion and shape of microliter-sized, liquid droplets.  While we derive a asymptotic, series solution for the shape of a droplet without van der Waals forces (\ref{AnalSol}), the solution could use a matching asymptotic solution near the back of the drop.  This would likely give insight into into the spread in data seen in this paper and others on droplet motion.  Also, asymptotics for small amounts of van der Waals forcing on this complete solution could complete the understanding of the long-time droplet shape changes.  Measurements of the precursor height, while not possible with current apparatus, could well become possible in the near future. This would perhaps allow close fitting of the dimensionless Hamaker constant and test the correct form for the van der Waals pressure term.

\section{\bf Appendix: Approximate Solution for the Droplet without van der Waals Forces}

  To understand the shape of the droplet we work in two dimensions, we turn to our change of variables (\ref{SimScale}).  Our governing equation (\ref{GovEqn}) under this rescaling becomes
\begin{equation}
\label{SGEA0}
	\phi_s + (\phi^3 - \xi \phi/3)_\xi = - (\phi^3 \phi_{\eta \eta \eta})_\eta + \mathcal{O}(e^{-2s/3}).
\end{equation}
  Long-time numerical simulations to this equation, show a single long time steady solution with gravitational forces $\phi^3$ balancing the term produced by the shrinking of $\xi$ scale in time (Fig.~\ref{Geometry} moved back to standard variables).  This solution is independent of initial condition shape as long as the initial droplet is reasonably concentrated (the initial width is less then half of the final transverse width).  This condition was not a concern as the initial droplet sizes in the experiments were significantly smaller than the final droplet in width.

  We start looking for an analytic, long-time form by removing the time dependence $\phi_s=0$.   For our analysis, we call $\xi = \xi_{\mathrm{tip}}$ to be the steady-state, tip position which will end up being close but not quite the actual tip position as we will ignore the capillary ridge seen in Figure~\ref{interf}.  We will define a new similarity position variable  $\bar{\xi}$ as the scaled distance up the drop from this point $\xi=\xi_{\mathrm{tip}}$.
\begin{equation*}
	\bar{\xi} = 1 - \frac{\xi}{\xi_{\mathrm{tip}}}
\end{equation*}

  In a study of fingering under gravity and surface tension gradients, Witelski and Jayaraman (private communication) introduce a mixed-spatial similarity variable they use to understand a balanced fourth-order smoothing perpendicular to their driving forces.  Since we also have a driving force perpendicular to our smoothing we use the same similarity variable $\bar{\eta}$ to modify our governing equation
\begin{eqnarray}
	\bar{\eta} = \frac{\eta}{(1-\xi/\xi_{\mathrm{tip}})^\alpha} = \frac{\eta}{\bar{\xi}^\alpha}.
\end{eqnarray}
 The similarity constant $\alpha$ takes a particular importance as the factor determining how the width of the droplet scales as a function of the distance from the tip (Fig.~\ref{footprint}).

  These new variables when put into the steady version of (\ref{SGEA0}) with $A=0$ produce the following equation for $\bar{\phi}(\bar{\xi},\bar{\eta}) = \phi(\xi,\eta)$
\begin{multline}
\label{tipPDE}
	\frac{1}{\bar{\xi}^{4\alpha}} (\bar{\phi}^3 \bar{\phi}_{\bar{\eta}\bar{\eta}\bar{\eta}})_{\bar{\eta}} + \frac{\alpha \bar{\eta}}{\xi_{\mathrm{tip}} \bar{\xi}} (\bar{\phi}^3)_{\bar{\eta}} - \frac{\alpha \bar{\eta}}{3 \bar{\xi}} (1 - \bar{\xi}) \bar{\phi}_{\bar{\eta}} \\
 = \pp{}{\bar{\xi}} \left( \frac{1}{\xi_{\mathrm{tip}}} (\bar{\phi}^3) - \frac{1- \bar{\xi}}{3} \bar{\phi} \right)
\end{multline}
 which appears quite daunting until we realize that we are interested in small $\bar{\xi} \ll 1$ near the tip of the drop.  In which case, we can set our scaling constant $\alpha = 1/4$ and look at the order $\mathcal{O}(1/\bar{\xi})$ equation
\begin{equation}
 	(\bar{\phi}^3 \bar{\phi}_{\bar{\eta}\bar{\eta}\bar{\eta}})_{\bar{\eta}} + \frac{1}{4 \xi_{\mathrm{tip}}} (\bar{\phi}^3)_{\bar{\eta}} - \frac{1}{12} \bar{\phi}_{\bar{\eta}} = 0
\end{equation}

  To find an approximate solution to this equation, we note that the problem is symmetric about the line $\eta=0$ and look for a series solution to the problem
\begin{equation}
	\bar{\phi}(\bar{\xi},\bar{\eta}) = f(\bar{\xi}) ( 1 + w_2 \bar{\eta}^2 + w_4 \bar{\eta}^4 + \ldots )
\end{equation}
 from which we find that after collecting terms of order $\bar{\eta}^0$ that $w_4$ must be $0$.  The full series solution for this equation does not seem to have an closed-form explicit solution.  However, since the equation skips three orders between the non-zero $w_2$ and $w_6$, we can hope to get a very reasonably accurate solution using only the first two terms of the expansion ($w_4=w_6=w_8=\ldots=0$).  Plugging in this shortened expansion into the full partial differential equation (\ref{tipPDE}) and collecting terms of order $\bar{\eta}^0$ we obtain just the terms from right hand side of that equation
\begin{equation}
	(f(\bar{\xi})^3)_{\bar{\xi}} =  \left( \frac{1-\bar{\xi}}{3} f(\bar{\xi}) \right)_{\bar{\xi}}.
\end{equation}

  This equation can be integrated once in $\xi$, but we must set the integration constant to zero to keep the solution from becoming infinite as $\xi \rightarrow 0$.  At this point we can just take the real solution of the cubic
\begin{equation}
	f(\bar{\xi}) = \sqrt{\frac{\xi_{\mathrm{tip}}(1-\bar{\xi})}{3}}.
\end{equation}
 and find the form for the solution along the center line $\eta=0$.

  To get a value for $w_2$ we would have to match this solution to an asymptotic solution near the base of the droplet (where $\xi \ll 1$) or perhaps multiple asymptotic solutions.  As it stands, efforts to produce such asymptotic solutions for the remainder of the droplet have not been successful.  However, we can leave $w_2$ as a fitting parameter like $\xi_{\mathrm{tip}}$ and fit solutions to the numerics and the experiments.  To make these comparisons, we move back into the scaled variables
\begin{equation}
\label{phiSolution}
	\phi(\xi,\eta;A=0) = \sqrt{\frac{\xi}{3}} \left( 1 -  \frac{\eta^2}{w^2} \right) + \mathcal{O}(y^6)
\end{equation}
where $w = W ( 1 - \xi/\xi_{\mathrm{tip}})^{1/4}$ and we have assumed from the numerics $w_2$ is negative and for convenience changed our fitting constant to $W = \sqrt{|w_2|}$. This series solution of \ref{SGEA0} (to $\mathcal{O}(y^2)$) is defined for $-w(\xi) < \eta < w(\xi)$ and $0 < \xi < \xi_{\mathrm{tip}}$ and the complete solution is defined to be zero elsewhere.  When this similarity solution is scaled back to the $h(x,y,t)$ variables using (\ref{SimScale}) we obtain the two-dimensional solution (\ref{AnalSol}) and its associated equation for the width of the droplet.

\section{Acknowledgments}

{\bf Acknowledgments:} We are grateful to support from the National Science Foundation under DMS grant \#0244498. We wish to thank Tom Witelski for careful reading and Bob Behringer and Alex Oron for helpful discussions.

\bibliographystyle{unsrt}
\bibliography{gravity}

\end{document}